\documentclass[11pt]{article}
\input epsf
\usepackage{amssymb, latexsym, amsmath}

\makeatletter \@addtoreset{figure}{section}
\def\thefigure{\thesection.\@arabic\c@figure}
\def\fps@figure{h, t}
\@addtoreset{table}{bsection}
\def\thetable{\thesection.\@arabic\c@table}
\def\fps@table{h, t}
\@addtoreset{equation}{section}

\newtheorem{theorem}{Theorem}[section]
\newtheorem{proposition}[theorem]{Proposition}
\newtheorem{lemma}[theorem]{Lemma}
\newfont{\tenbi}{cmbxti10}

\def\la {\lambda}

\begin{document}
\title{Closed Geodesics and Billiards on Quadrics related to elliptic KdV
solutions
\footnote{AMS Subject Classification 37J60, 37J35, 70H45}}

\author {Simonetta Abenda \\
Dipartimento di Matematica e CIRAM \\
Universit\`a degli Studi di Bologna, Italy \\
{\footnotesize  abenda@ciram.unibo.it} \\
and \\
Yuri Fedorov \\
Department of Mathematics and Mechanics
\\ Moscow Lomonosov University, Moscow, 119 899, Russia \\
{\footnotesize  fedorov@mech.math.msu.su} \\ and \\
Departament de Matematica I \\
Universitat Politecnica de Catalunya, Spain \\
{\footnotesize Yuri.Fedorov@upc.es} }
\maketitle
\begin{abstract}
We consider algebraic geometrical properties of the integrable billiard on
a quadric $Q$ with elastic impacts along another quadric confocal to $Q$.
These properties are in sharp contrast with those of the ellipsoidal Birkhoff
billiards in ${\mathbb R}^n$. Namely,
generic complex invariant manifolds are not Abelian varieties,
and the billiard map is no more algebraic. A Poncelet-like
theorem for such system is known. We give explicit sufficient
conditions both for closed geodesics and periodic billiard
orbits on $Q$ and discuss their relation with the elliptic
KdV solutions and elliptic Calogero system.
\end{abstract}

\section{Introduction}
One of the best known discrete integrable systems is the billiard inside an
$n$-dimensional ellipsoid (more generally, a quadric)
\begin{gather*}
Q = \left\{ \frac{X_1^2}{a_1}+\cdots +\frac{X_{n+1}^2}{a_{n+1}}= 1
\right\}\in {\mathbb R}^{n+1}, \qquad
{\mathbb R}^n=(  X_1,X_2,\dots,X_{n+1}), \\
0 < a_1 < \cdots < a_n< a_{ n + 1}
\end{gather*}
with elastic refections on $Q$\cite{Birk}. The billiard inside $Q$
(Birkhoff case) inherits the remarkable property of geodesics on
$Q$ given by the Chasles theorem: the trajectories or their
continuations before and after impacts are tangent to the same set
of $n$ quadrics that are confocal to $Q$ (see, e.g.,
\cite{Moser}). The parameters of these quadrics deliver $n$
independent and commuting first integrals of the discrete system.

Veselov \cite{2} described the Birkhoff billiard map in terms of
discrete Lagrangian formalism and showed that its complex
invariant manifolds are open subsets of coverings of Jacobians of
hyperelliptic curves. Moreover,  the restriction of the map to
such manifolds is represented by shifts by a constant vector,
which gives rise to an addition law on Abelian varieties.

Explicit theta function solutions for this billiard were obtained
in \cite{2} by applying spectral theory of difference operators,
and later in \cite{Fed1} by finding a degeneration of theta
function solutions for geodesics on an ellipsoid of a higher
dimension.

The periodicity problem for the Birkhoff billiard map is closely
related to a generalization of the following geometrical problem:
given two smooth conics in the projective plane, $C$ and $D$, to
construct a closed polygon inscribed in $C$ and circumscribed
about $D$. The full Poncelet theorem states that, if a closed
polygon exists for given $C$ and $D$, then every point of $C$ is a
vertex of a closed polygon and all such polygons have the same
number of sides \cite{GH2}.

Cayley theorem gives an explicit condition to find such a polygon.
Namely, the iterative process of constructing the polygonal line
is equivalent to shifts by a constant vector $\mathbf v$ in the
Jacobian of an elliptic curve, $\cal E$. Then the polygonal line
is closed after $n$ steps if and only if $\mathbf v$ is neutral on
$\cal E$ (\cite{Cayley, GH1}).

 The relation of the Poncelet problem to
periodic billiard orbits inside $Q$ is evident, and such orbits
are described by generalizations of the theorems of Poncelet and
of Cayley in \cite{Chang_Fri, Chang0, Drag_Rad, EP1}.

\smallskip

On the other hand, one can consider geodesic motion on the
ellipsoid $Q$ combined with elastic reflections along its
intersections with the confocal quadric
$$
Q_d= \left\{ \frac{X_1^2}{a_1-d}+\cdots
+\frac{X_{n+1}^2}{a_{n+1}-d}=1 \right\},
$$
$d$ being an arbitrary fixed parameter. Thus one obtains the
billiard on the ellipsoid described by the map ${\cal B}\, :
(x,v)\mapsto (\tilde x, \tilde v)$, where $x,v\in {\mathbb
R}^{n+1}$ are respectively the coordinates of an impact point on
$Q\cap Q_d$ and the outgoing velocity at this point, whereas
$\tilde x, \tilde v$ denote the same objects at the next impact
point.

As shown in \cite{Chang}, this billiard map is also integrable and
the trajectories between impacts (arches of geodesic) or their
continuations are tangent to the same set of $n-1$ quadrics
(caustics) in ${\mathbb R}^{n+1}$ that are confocal to $Q$ and
$Q_d$. For $n=2$, algebraic geometrical features of this problem
have already been discussed in \cite{Ves2}.

Next, following  \cite{Chang}, if a billiard trajectory of $Q$ is
closed, then all the trajectories sharing the same caustics are
closed as well and have the same period and length. Thus one
arrives at a generalization of the celebrated Poncelet theorem on
a quadratic surface.

\paragraph{Contents of the paper.}
In this paper paper we study algebraic geometrical properties and
periodic orbits of the billiard on the ellipsoid $Q$ with impacts
along its intersection with a confocal quadric $Q_d$.

All our statements in the complex setting also hold in the case of
a general quadric $Q$. On the other hand, in the description of
real closed geodesics and periodic billiard orbits we always
assume $Q$ to be an ellipsoid, since, for a generic real quadric
$Q$, the billiard problem becomes meaningful only after a
compactification.

We show that the algebraic-geometrical properties of the billiards
on $Q$ with impacts along $Q_d$ are in sharp contrast with those
of the Birkhoff case. Namely, the generic complex invariant
manifolds are not Abelian varieties, but open subsets of
theta-divisors of $n$-dimensional hyperelliptic Jacobians.

Next, although the billiard map $\cal B$ is integrable and all of
its first integrals are algebraic, in the complex coordinates
$(x,v)$ it is infinitely many valued, whereas in the real domain
the new values $(\tilde x, \tilde v)$ are determined by solving a
system of transcendental equations which involve the inversion of
hyperelliptic integrals.

To our knowledge, such billiards actually provide a first
nontrivial example of a discrete integrable system which cannot be
described in terms of an algebraic addition law on Abelian varieties.

Such systems can be regarded as discrete analogs of
hyperelliptically separable systems, a class of finite-dimensional
integrable differential equations whose generic complex solutions
have movable algebraic branch points and are single-valued on an
{\it infinitely} sheeted ramified covering of the complex plane
$t$ (see, e.g., \cite{Van, AF, AlberFed, ERP}).
\medskip

In Section 2 we briefly recall an algebraic geometrical
description and theta-function solution of the Jacobi problem on
geodesics on a quadric $Q$, then describe the main properties of
the billiard on $Q$.

In Section 3, we consider the periodicity problem for the billiard
map ${\cal B}$ and we show that, in general, such conditions
cannot be formulated in an algebraic form.

This however becomes possible for special initial conditions for
which the corresponding hyperelliptic curve $\Gamma$ is a covering
of an elliptic curve $\cal E$ and one of the holomorphic
differential on $\Gamma$ reduces to that on $\cal E$. Under these
conditions, the corresponding geodesics on $Q$ have a finite
number of intersections with the boundary $Q \cap Q_d$ and
therefore are closed.

In Section 4, we use known properties of the periodic KdV
solutions and of the solutions to the elliptic Calogero
system\cite{Dub_Nov,Cal1,AMM77,Krich,Smirnov1,Smirnov2,Gavr_Per,Tr}
to describe the curves $\Gamma$ and the related closed geodesics
on $Q$. As a result we show that to each elliptic $N$-soliton KdV
solution or solution of the $N$-body elliptic Calogero systems
satisfying the locus condition there corresponds a family of
closed geodesics on a quadric $Q$.

We consider in detail the case of the 3:1 coverings $\Gamma\mapsto
{\cal E}$ and obtain explicit sufficient conditions for a real
geodesic on a 2-dimensional ellipsoid $Q$ to be closed.

Finally, in Section 5 we derive sufficient conditions for the
periodicity of the billiard map on $Q$ by describing admissible
boundary parameters $d$ of $Q_d$. To our knowledge, this is the
first time a sufficient condition of periodicity for such billiard
map is given.

In the conclusion we outline possible extensions of this approach
to periodic billiard orbits on a multi-dimensional ellipsoid $Q$.

\section{Geodesics and Billiards on a Quadric}
We start with the celebrated Jacobi problem of the geodesic motion
on an $n$-dimensional ellipsoid $Q$. The problem is well known to
be integrable and to be linearized on a covering of the Jacobian
of a genus $n$ hyperelliptic curve. Namely, let $l$ be the natural
parameter of the geodesic and $\la_1,\dots,\la_n$ be the
ellipsoidal coordinates on $Q$ defined by the formulas
\begin{equation}
\label{spheroconic2}
X_i=\sqrt{ \frac{(a_i-{\la}_1)\cdots (a_i-{\la}_n)}
{ \prod_{j\ne i}(a_i-a_j)} }, \qquad i=1,\dots, n+1 .
\end{equation}
Then, denoting $\dot\lambda_k= d\lambda_k/d l$ the corresponding
velocities, the total energy $\displaystyle \frac 12 (\dot X, \dot
X)$ takes the St\"ackel form
\begin{gather*}
H = 2\sum^{n}_{k=1} \frac 1{\Psi (\lambda_{k})} \prod\limits^{n}_{j\ne k}
(\lambda_{k}-\lambda_{j}) \, {\dot\lambda}^{2}_{k}\, , \qquad
\Psi (\lambda)=(\lambda-a_{1})\cdots(\lambda-a_{n+1}).
\end{gather*}
According to the St\"ackel theorem, the system is Liouville
integrable and, after re-parametrization
\begin{equation} \label{tau-1}
dl =\la_1\cdots\la_n\, ds ,
\end{equation}
the evolution of $\lambda_{i}$ is described by quadratures
\begin{gather}
\frac{\la_1^{k-1} d\la_1}{2\sqrt{- \Psi(\la_1) {R} (\la_1) }} +\cdots
+ \frac{\la_n^{k-1} d\la_n}{2\sqrt{-\Psi(\la_n) {R}(\la_n)}}
 =\Bigg \{
\begin{aligned}
ds\quad \mbox{ for } & k=1\, , \\
0\quad \mbox{ for } & k=2, \dots, n ,
\end{aligned} \label{quad2} \\
{R}(\la) =\la (\la-c_1)\cdots (\la-c_{n-1}), \nonumber
\end{gather}
$c_k$ being constants of motion. Then the components of the velocity
have the form
\begin{equation} \label{spheroconic3}
V_i=\frac{d X_{i}}{dl} = {\sqrt{a_i (a_i -\lambda_i )\cdots (a_i
-\lambda_{n})} \over \sqrt{\prod_{j\ne i}(a_i-a_j)} }
\sum^{n}_{k=1} {1\over \prod\limits_{j\ne k}
(\lambda_{k}-\lambda_{j})} {\sqrt{-\Psi(\lambda_k) R(\lambda_k)}
\over \lambda_{k}(a_{i}-\lambda_{k})}\,  .
\end{equation}

The quadratures involve $n$ independent holomorphic differentials
on the genus $n$ hyperelliptic curve
$\Gamma=\{\mu^2=-\Psi(\la){R}(\la)\}$ and give rise to the
Abel--Jacobi map of the $n$-th symmetric product $\Gamma^{(n)}$ to
the Jacobian variety of $\Gamma$,
\begin{gather}
\int \limits^{P_1}_{P_0} \omega_k + \cdots + \int
\limits^{P_n}_{P_0} \omega_k =u_{k}, \qquad
k=1,\dots, n, \label{AB} \\
\omega_k =\frac{\la^{k-1} d\la }{2 \sqrt{- \Psi(\la) {R} (\la) }}\, , \quad
P_k=\left(\lambda_k,\sqrt{- \Psi(\la_k) {R} (\la_k)} \right ) \in \Gamma, \nonumber
\end{gather}
where $u_1,\dots,u_n$ are coordinates on the universal covering of
Jac$(\Gamma)$ and $P_0$ is a fixed basepoint, which we choose to
be the infinity point, $\infty$, on $\Gamma$.

Since $u_1=s +$const and $u_2,\dots,u_n=$const, the geodesic
motion in the new parametrization is linearized on the Jacobian
variety of $\Gamma$.

For the classical Jacobi problem (n=2), its complete
theta-functional solution was presented in \cite{Weier}, and, for
arbitrary dimensions, in \cite{Knorr}, whereas a complete
classification of real geodesics was made in \cite{Audin}.

Namely, let us fix a canonical homology basis on $\Gamma$ and let
$\bar\omega_1,\dots,\bar\omega_n$ be the conjugated basis of
normalized holomorphic differentials,
\begin{equation}\label{coordchange}
\bar\omega_i= \sum_{k=1}^n C_{ik} \, \omega_k\, ,
\end{equation}
$C_{ik}$ being normalization constants so that the $2n\times n$
period matrix of $\Gamma$ has the form $(2\pi \jmath\, {\bf I},\;
B)$, where $B$ is the $n\times n$ Riemann matrix.

Let $\varphi=(\varphi_1,\dots,\varphi_n)^T$, $\varphi_i= \sum_{s=1}^n C_{is} u_s$
be the corresponding local coordinates on Jac$(\Gamma)$ and
\begin{gather} \label{char}
\theta\! \left[{ \alpha \atop \beta}\right]\! (\varphi|B) =\exp
\{\langle B\alpha ,\alpha \rangle /2+\langle \varphi +2\pi
\jmath\beta,\alpha \rangle\}\;
\theta (\varphi +2\pi \jmath\beta+B\alpha) , \\
\jmath=\sqrt{-1} \nonumber
\end{gather}
be associated theta-function with characteristics $\alpha
=(\alpha_{1},\ldots,\alpha_{n})$, $\beta
=(\beta_{1},\ldots,\beta_{n})$ and  for $K,M\in {\mathbb Z}^{n}$,
\begin{gather}
\theta\! \left[ { \alpha \atop \beta } \right ] \! (\varphi +2\pi \jmath K+BM)
=\exp (2\pi \jmath\, \epsilon)\exp \{-(BM,M)/2-(M,\varphi)\}
\theta\! \left[ { \alpha \atop \beta } \right ]  \!(\varphi) , \label{1.5} \\
\epsilon =(\alpha ,K)-(\beta ,M)\, . \nonumber
\end{gather}
Then the inversion of the map (\ref{AB}) applied to formulas
(\ref{spheroconic2}) leads to the following parametrization of a
generic geodesic
%in terms of $n$-dimensional theta-functions associated to the curve $\Gamma$
%\footnote{Here we assume that the arguments of $\theta$ are associated to the canonical
%basis of holomorphic differentials $\{\omega_i\}$, not normalized ones.}
\begin{gather} \label{theta_n}
X_i (s ) = \varkappa_i \frac{ \theta[\Delta+\eta_i] (U_1 s +\varphi_0) }
{\theta[\Delta] (U_1 s  +\varphi_0) }, \qquad i=1,\dots, n+1, \\
U= 2(C_{11},\dots, C_{n1}) . \nonumber
\end{gather}
Here $\Delta=(\delta'',\delta')$,
$\eta_{i}=(\eta_{i}'',\eta_{i}')\in {\mathbb R}^{2n}/2 {\mathbb
R}^{2n}$ are half-integer theta-characteristics such that modulo
the period lattice of $\Gamma$ the following relations hold
\begin{equation}
\label{etas}
2\pi \jmath\, \eta_{i}'' +B \eta_{i}'
=\int_{\infty}^{(a_i,0)} (\bar\omega_1,\dots,\bar\omega_n)^T, \quad
2\pi \jmath \delta'' +B \delta'={\cal K} ,
\end{equation}
$\cal K$ being the vector of the Riemann constants, and
$\kappa_i$ are constant factors depending on the moduli of $\Gamma$ only.

\paragraph{Remark 2.1.}
As follows from the quasiperiodic law (\ref{1.5}), expressions (\ref{theta_n}) are
meromorphic functions on a covering $\widetilde{\rm Jac}(\Gamma)$
of Jac($\Gamma$) obtained by doubling some of its periods (see also \cite{Knorr}).

\paragraph{The billiard on $Q$.} We now describe the billiard map
associated to the geodesic motion of a point on $Q$ with elastic
bounces along the confocal quadric $Q_d$. By (\ref{spheroconic2}),
when $(X_1,\dots,X_{n+1})\in Q\cap Q_d$, one of the points
$P_i=(\lambda_i,\mu_i)$ on $\Gamma$ (without loss of generality we
choose it to be $P_n$) coincides with one of the points $E_{d
\pm}=(d,\pm \sqrt{R(d)})$. Under the Abel--Jacobi map (\ref{AB}),
the condition $P_n=E_{d\pm}$  defines two codimension one
subvarieties in ${\rm Jac} (\Gamma)$, the theta-divisors
$$
\Theta_{d\pm} =\{ \theta[\Delta](\varphi \mp C{\mathfrak q}/2) \}
\in {\rm Jac}(\Gamma),
$$
where $C=(C_{ik})$ is the matrix introduced in (\ref{coordchange})
and $\mathfrak q$ is the vector
\begin{equation} \label{qq}
{\mathfrak q} =({\mathfrak q}_1,\dots,{\mathfrak q}_n)^T
= \int_{E_{d-}}^{E_{d+}} (\omega_1,\dots,\omega_n)^T \in {\mathbb C}^{n}.
\end{equation}
$\Theta_{d+}$ and $\Theta_{d-}$ have a common $(n-2)$-dimensional
subvariety.

Hence, for fixed constants of motion, the coordinates $x$ of
impact points on $Q\cap Q_d$ with velocities $v$ are described by
degree $n-1$ divisors $\{P_1,\dots,P_{n-1}\}$, that is, by a point
$\varphi$ on $\Theta_{d-}$ or $\Theta_{d+}$.
\medskip

An algebraic geometric description of the motion between impacts
and at the elastic bounce along $Q\cap Q_d$ is given by the
following proposition.

\begin{proposition} \label{main0}
The geodesic motion on $Q$ between subsequent impact points $x$
and $\tilde x$ on $Q\cap Q_d$ corresponds to the straight line
uniform motion on ${\rm Jac} (\Gamma)$ between $\Theta_{d+}$ and
$\Theta_{d-}$ along the $u_1$-direction. The point of intersection
with $\Theta_{d-}$ gives the next coordinate $\tilde x$ and the
ingoing velocity $v'$. At the point $\tilde x$ the reflection $v'
\to \tilde v$ (from the ingoing to the outgoing velocity) results
in jumping back from  $\Theta_{d-}$ to $\Theta_{d+}$ by the shift
vector $\mathfrak q$, which does not change $\tilde x$.  Then the
procedure iterates.
\end{proposition}

The process is sketched in Figure 1.
\medskip

\epsfysize=6cm
\epsfxsize=10cm
\epsfbox{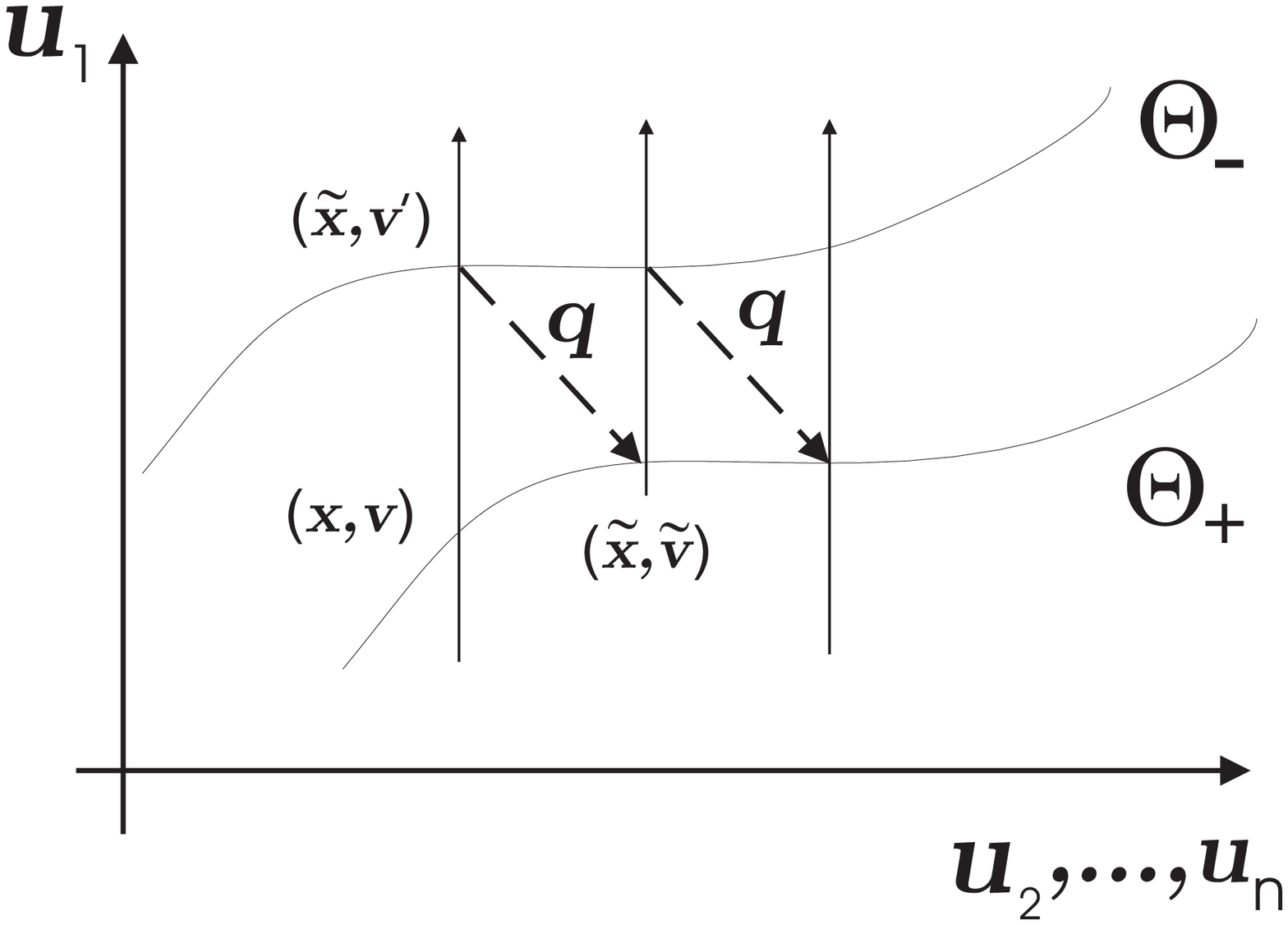}

\centerline{Figure 1}

\bigskip

\noindent{\it Proof of Proposition} \ref{main0}. First, we want to
show that any subsequent impact points and the velocities $(x,v)$
and $(\tilde x, v')$ correspond to intersections of the flow with
$\Theta_{d+}$ and $\Theta_{d-}$ respectively. For this purpose, we
introduce the new time parameter $\Xi$ such that
$$
d\Xi =\frac {2 \sqrt{- \Psi(d) R(d)} } {(\la_1-d)\cdots (\la_n-d)}\, ds \, .
$$
Notice that in the real case the number $\sqrt{- \Psi(d) R(d)}$ is
real and positive. Hence, as the point on $Q$ moves from $x$ to
$\tilde x$, the parameter $\Xi$ changes monotonically from
$-\infty$ to $\infty$. On the other hand, in view of the
quadratures (\ref{quad2}), one has
$$
d\Xi = \sum_{j=1}^n
\frac{\sqrt{- \Psi(d) R(d)} \, d\la_j}{(\la_i -d)\sqrt{- \Psi(\la_j) {R}(\la_j) }}
$$
and, therefore,
\begin{equation} \label{AbTr}
\Xi= \sum_{j=1}^n \int_{P_0}^{P_j} \frac{\sqrt{- \Psi(d) R(d)}
d\la}{(\la -d)\sqrt{ -\Psi(\la_j) {R} (\la_j)}}+\mbox{const} .
\end{equation}
The latter expression contains a meromorphic differential of the
third kind on the curve $\Gamma$ having the pair of simple poles
at $E_\pm=(d,\pm \sqrt{- \Psi(d) R(d)} )$ with residue $\pm 1$
respectively, hence the sum (\ref{AbTr}) is an Abelian
transcendent of the third kind. According to the theory of
theta-functions (see e.g., \cite{ClGor} or \cite{BEBIM}), under
the Abel--Jacobi map (\ref{AB}) the sum takes the form
$$
\Xi = \log \frac {\theta[\Delta](\varphi +C{\mathfrak q}/2)}
{\theta[\Delta](\varphi -C{\mathfrak q}/2)} + {\cal C}s  + \mbox{const},
$$
where $\cal C$ is a normalizing factor, $C=(C_{ik})$ is defined in
(\ref{coordchange}) and the vector $\mathfrak q$ is defined in
(\ref{qq}). It follows that for $\Xi=-\infty$ and $\Xi =\infty$
the point $\varphi$ in Jac$(\Gamma)$ belongs to $\Theta_{d+}$ and
$\Theta_{d-}$ respectively, which proves the first statement of
the theorem.

Next, we notice that for any $\varphi'\in \Theta_-$ and $\tilde
\varphi\in\Theta_+$ such that $\tilde \varphi=\varphi'+C{\mathfrak
q}$, the formula (\ref{theta_n}) gives one and the same Cartesian
coordinates $X_i$. Indeed, under the mapping (\ref{AB}), $\tilde
\varphi$ and $\varphi'$ are represented by the degree $n$ divisors
$$
\{P_1=(\lambda_1,\mu_1),\; \dots,\; P_{n-1}=(\lambda_{n-1},\mu_{n-1}),\, E_-\}
$$
and, respectively, $\{P_1,\dots, P_{n-1}, E_+\}$, which by
(\ref{spheroconic2}), lead to the same values of $X_i$. On the
other hand, expressions (\ref{spheroconic3}) imply that for
$\tilde \varphi$ and $\varphi'$ the velocities $V_i$ are
different.

As a result, the intersection point $\varphi'\in \Theta_-$
corresponds to the pair $(\tilde x, v')$, whereas $\tilde
\varphi=\varphi'+C{\mathfrak q}\in \Theta_+$ gives rise to $\tilde
x$ and the outgoing velocity $\tilde v$. \boxed{}
\medskip

It follows that the restriction of the billiard map $\mathcal B\,
: \, (x,v)\to (\tilde x, \tilde v)$ onto its invariant tori can be
regarded as a transformation ${\mathcal B}_c \; :\,  \Theta_{d+}
\to \Theta_{d+}$. Then, as a corollary of Proposition \ref{main0},
we obtain

\begin{theorem}  \label{main1}
The map  ${\mathcal B}_c$ is given by the shift of coordinates
$u_2,\dots, u_{n}$ by ${\mathfrak q}_2,\dots,{\mathfrak q}_{n}$
respectively, whereas the first coordinate $u_1$ on ${\rm
Jac}(\Gamma)$ is chosen such that the vector $\tilde u={\mathcal
B}_c(u)$ remains on the theta-divisor $\Theta_{d+}$.
\end{theorem}

Indeed, according to Proposition \ref{main0} (we recall that
$\varphi =Cu$), under the straight line flow on the Jacobian from
$\Theta_{d+}$ to $\Theta_{d-}$, the coordinates $u_2,\dots, u_{n}$
remain unchanged, and under the passage to $\Theta_{d+}$ they
increase by ${\mathfrak q}_2,\dots, {\mathfrak q}_{n}$.

\paragraph{Remark 2.2.}
For $n=2$, the above algebraic geometrical property has already
been described in \cite{Ves2}. The algebraic geometric description
of the billiard on $Q$ stands in sharp contrast to the one of the
Birkhoff billiard system - as well as to most of the known
integrable maps. Indeed in the Birkhoff case, the billiard map is
given by a shift by a constant vector on a complex invariant
manifold which is an open subset of an Abelian variety. In our
case, the shift takes place on a non-Abelian variety and, as a
result, it is {\it not fixed}.
\medskip

\paragraph{Branching of the map} ${\mathcal B}_c$.
The above properties of the map $\cal B$ cannot be used to derive
an explicit expression for the billiard map $(x,v)\to (\tilde
x,\tilde v)$, since the complex map ${\mathcal B}_c \; :\,
\Theta_{d+} \to \Theta_{d+}$ described by Theorem \ref{main1} is
infinitely many-valued. Indeed, let $\widetilde\Theta_{d+}$ be the
universal covering of $\Theta_{d+}$ in ${\mathbb C}^n=(u_1,\dots,
u_{n})$ and let $\pi$ denote the natural projection
$\widetilde\Theta_{d+} \to {\mathbb C}^{n-1} =(u_2,\dots, u_{n})$.

\begin{theorem} \textup{(see also \cite{AF})}. Suppose that the
curve $\Gamma$ does not have nontrivial symmetries and its
Jacobian does not possesses Abelian subvarieties, then
\begin{description}
\item{1).} Under the  projection $\pi$, $\widetilde\Theta_{d+}$ is
an infinitely sheeted covering of ${\mathbb C}^{n-1}$ ramified
along a codimension one subvariety of $\widetilde\Theta_{d+}$.
\item{2).} A complete preimage of any point in ${\mathbb C}^{n-1}$
consists of an infinite number of points on
$\widetilde\Theta_{d+}$, which, in turn, represent an infinite
number of points on $\Theta_{d+}$.
\end{description}
\end{theorem}

\noindent{\it Proof.} Let ${\mathfrak v}_1,\dots,{\mathfrak
v}_{2n}\in {\mathbb C}^{n-1} =(u_2,\dots,u_n)$ be the projections
of $2n$ independent period vectors of Jac$(\Gamma)$. For any point
${\mathcal P}\in \Theta_{d+}$, the projections of points of its
equivalence class on $\widetilde\Theta_{d+}$ have the form
$$
\bigg \{ \pi({\mathcal P})+\sum_{j=1}^{2n} m_j {\mathfrak v}_{j}
\,|\, m_j\in {\mathbb Z} \bigg \}.
$$
Let $\hat z$ be any fixed point on ${\mathbb C}^{n-1}$.
Under the condition of the theorem, the projections ${\mathfrak v}_{j}$ are
incommensurable and the integer coefficients
$m_j$ may always be chosen in such a way that the above sum represents a
point in any small and a priori defined neighborhood of $\hat z$. In this sense,
any point on $\Theta_+$ is projected into such a neighborhood of any point on
${\mathbb C}^{n-1}$.
Hence, a complete preimage $\pi^{-1}$ of any point on ${\mathbb C}^{n-1}$
corresponds to an infinite number of points on the theta-divisor.
\medskip

In the particular case $n=2$, when the theta-divisor
or its translate coincides with the genus 2 curve $\Gamma$ itself, the above
property is known in connection with the inversion of a single hyperelliptic integral
$$
\int_{(\lambda_0, w_0)}^{(\lambda, w)}\frac{d\lambda}w =u_1,
\qquad (\lambda, w)\in \Gamma.
$$
Namely, as was shown by Jacobi (see, e.g., \cite{Mark}), the
covering $\widetilde\Gamma\to {\mathbb C}=\{u_1\}$ has infinitely
many branch points whose projections onto  the complex plane $u_1$
form  a dense set.

Hence, given fixed coordinates $u_2,\dots, u_{n}$ on
Jac$(\Gamma)$, there is an infinite number of $\tilde u_1$, which
define the next impact point ${\tilde x}\in Q \cap Q_d$ via
Proposition \ref{main0}. In particular, this implies that the
billiard map ${\mathcal B}\; :\,(x,v)\to ({\tilde x}, {\tilde v})$
cannot be algebraic.

The above phenomenon is also explained by the fact that a generic
geodesic on $Q$ intersects any connected component of the boundary
$Q \cap Q_d$ at infinitely many points. In the real domain, given
$x\in Q \cap Q_d$, it is possible to determine $\tilde x$ by
choosing the {\it next} intersection with the boundary, but in the
complex setting such a choice cannot be made.
\medskip

\paragraph{Remark 2.3.}
Due to the above properties, the billiard map $\cal B$ can be
regarded as a discrete analog of the hyperelliptically separable
systems studied, in particular, in \cite{Van,AF,AlberFed,ERP}).
Such systems, although being Liouville integrable, are not
algebraic completely integrable: their generic complex solutions
have movable algebraic branch points and are single-valued on an
{\it infinitely} sheeted ramified covering of the complex plane
$t$, so they possess the so called weak Painlev\'e property. The
branching of the billiard map $\cal B$ can be viewed as a discrete
version of such a property (see also \cite{Gramm}).

\section{Periodic Billiard Orbits on $Q\cap Q_d$}
% and a Generalized Poncelet Theorem.} ???
We now concentrate on algebraic conditions for
a billiard trajectory on the quadratic surface $Q$ to be periodic.

As mentioned in the Introduction, the Birkhoff billiard in the
ellipsoid $Q\subset {\mathbb R}^{n+1}$ is described by
translations on the Jacobians of genus $n$ hyperelliptic curves
$$
\Gamma=\{y^2 =-R(x) \}\subset {\mathbb R}^2=(x,y), \quad R(x)=(x-b_1)\cdots (x-b_{2n+1}).
$$
Then the corresponding billiard trajectory is periodic if, for
a certain integer $m$, the vector
$$
m\,\int^{E_{0+}}_{E_{0-}}(\omega_1,\dots,\omega_n)^T, \qquad
E_{0\pm}=(0,\pm\sqrt{-R(0)})
$$
is neutral in Jac$(\Gamma)$. The periodicity conditions written
explicitly in terms of the coefficients of the curve $\Gamma$ were
obtained in \cite{Drag_Rad} by proving the following theorem.

\begin{theorem} \label{Drag0} \textup{(\cite{Drag_Rad})}
For a constant $d\in {\mathbb C}$ and involutive points $E_{d\pm }
=(d,\pm\sqrt{R(d)})\in \Gamma$, the vector
$m\,\int^{E_{d+}}_{E_{d-}}\, \omega$ ($m>n$) is neutral in
\textup{Jac}$(\Gamma)$ if and only if
\begin{equation} \label{rank}
\textup{rank } \begin{pmatrix} S_{m+1} & S_m \cdots S_{n} \\
                              S_{m+2}  & S_{m+1}  \cdots S_{n+1} \\
                               \vdots  & \vdots  \ddots \vdots \\
 S_{2m-1}  & S_{2m-2}  \cdots S_{m+n}  \end{pmatrix} < m-n,
\end{equation}
where the entries $S_j$ are determined from the expansion
$$
\sqrt {(x-b_1)\cdots (x-b_{2n+1})} =S_0 + S_1 (x-d) +S_2(x-d)^2 +\cdots .
$$
\end{theorem}
This theorem provides a multidimensional generalization of
Cayley's algebraic condition for the existence of a polygon
inscribed in a conic and circumscribed about another conic.
\medskip

For the billiard {\it on} the quadric $Q$, the situation is quite
different: generic complex invariant manifolds are no more Abelian
varieties, but theta-divisors. As follows from Theorem
\ref{main1}, the billiard trajectory on $Q$ is periodic if for
some integers $m, m_1,\dots, m_{2n}$ the following equality holds
\begin{equation} \label{compare}
m ({\mathfrak q}_2,\dots, {\mathfrak q}_{n})^T =\sum_{j=1}^{2n} m_j {\mathfrak v}_{j} ,
\end{equation}
Since the projections are generally incommensurable, the above
condition can always be fulfilled, hence {\it in the complex
setting} this criterion for closed billiard trajectories makes no
sense.
\medskip

\paragraph{Remark 3.1.}
On the other hand, it seems natural to try to replace the above
condition with the following relation in the Jacobian:
$$
 m \int_{E_{d-}}^{E_{d+}} (\omega_1,\dots,\omega_n)^T \equiv 0 .
$$
That is, although the map ${\cal B}_c$ is not given by a shift by
$\mathfrak q$ on Jac$(\Gamma)$, it is required that after $N$
iterations it yields an identity map on Jac$(\Gamma)$ and
therefore, on the theta-divisor $\Theta_{d+}$ as well. This
condition was proposed, amongst other results, in the recent paper
\cite{Drag_Rad_3}, and explicit algebraic conditions on the
coefficients of the curve $\Gamma$ were derived from it. However,
since a generic geodesic $X(s )\subset Q$ intersects $Q \cap Q_d$
at infinitely many points, these conditions do not guarantee that
the corresponding closed billiard trajectory consists of pieces of
geodesics joining the {\it nearest} points of intersection with $Q
\cap Q_d$.

\paragraph{Periodic billiard orbits related to closed geodesics.}
The projections ${\mathfrak v}_1,\dots,{\mathfrak v}_{2n}$  may
become commensurable and the closeness condition in the complex
domain may become meaningful if the hyperelliptic curve $\Gamma$
has certain symmetries and its Jacobian contains Abelian
subvarieties, as described by the Weierstrass--Poincar\'e theory
of reduction of Abelian varieties and functions (see e.g.,
\cite{Kraz} and, for modern exposition, \cite{BEBIM, Enol}).

In this paper we concentrate on the case when the genus $n$
hyperelliptic curve $\Gamma_n$ is an $N$-fold covering of an
elliptic curve $\cal E$ and the ``first'' holomorphic differential
$\omega_1=d\lambda /\mu$ reduces to a holomorphic differential on
$\cal E$.

Under these conditions, the $u_1$-flow on Jac$(\Gamma)$ intersects $\Theta_{d+}$
at a finite number of points. As a result,
the corresponding geodesics on $Q$ also have a finite number of intersections with
the boundary $Q \cap Q_d$ and therefore are closed.

According to the Poincar\'e theorem, the existence of covering
$\Gamma \mapsto {\cal E}$ implies certain conditions on the
components of the Riemann matrix of the curve $\Gamma$
(\cite{BEBIM, Enol}). However, our objective below is to give
sufficient conditions in terms of the coefficients of the
polynomial $R(\lambda)$.

\paragraph{Remark 3.2.} We emphasize that not any covering
$\Gamma \mapsto \cal E$ corresponds to closed geodesics. For
example, in the simplest possible case $n=2$, $N=2$ studied by
Jacobi and widely described in the literature (see, e.g.
\cite{Kraz, BEBIM}), the curve $\Gamma$ is birationally equivalent
to the canonical curve
$$
w^2 = z(z-1)(z-\alpha)(z-\beta)(z-\alpha \beta),
$$
$\alpha, \beta$ being arbitrary constant. It covers two elliptic curves
$$
\ell_\pm = \{W_\pm^2=Z_\pm (1-Z_\pm)(1-k_\pm Z_\pm) \}, \qquad
k_\pm^2=- \frac {(\sqrt{\alpha}\mp \sqrt{\beta} )^2}{(1-\alpha)(1-\beta)}
$$
and both of the holomorphic differentials $\omega_1, \omega_2$ on
$\Gamma$ reduce to linear combinations of the holomorphic
differentials on $\ell_+$ and $\ell_-$.

In this case the inversion of the quadratures (\ref{quad2}) and
formulas (\ref{theta_n}) lead to solutions in terms of elliptic
functions of $\ell_\pm$, but, since their periods are generally
incommensurable, the corresponding geodesics are quasi-periodic.
Such geodesics were already studied in \cite{Braun}.

\section{Elliptic KdV solutions, elliptic Calogero systems and closed geodesics
on a quadric}

The most classical series of {\it hyperelliptic tangential}
coverings  ${\cal G}_n \to {\cal E}$ (see \cite{TV}) arises in the
spectral theory of the Lam\'e potentials or as a subclass of
spectral curves of the elliptic Calogero (EC) systems, which
describe the motion of $N$ particles with coordinates $q_1,\dots,
q_N$ on the circle under the potential $ \sum_{i<j}^N \wp
(q_i-q_j)$, see \cite{Cal1}. Here $\wp(q)=\wp (q\mid
\omega,\omega')$ denotes the Weierstrass elliptic function with
half-periods $\omega,\omega'$.

The EC systems are integrable and, as shown in \cite{Krich}, they
possess a matrix $N\times N$ Lax pair with an elliptic spectral
parameter $u \in {\cal E}={\mathbb C}/\{ 2\omega{\mathbb Z}+
2\omega'{\mathbb Z}\}$ \footnote{We do not give the explicit
expressions for $L(u)$ here},
$$
\frac d{dt} L(u) =[L(u), M(u)].
$$
The spectral curve ${\cal G}=|L(u)-\mu {\bf I}|=0$ (the so called
Krichever curve) is an $N$-fold covering of the torus $\cal E$ and
its genus does not exceed $N$.

The EC systems possess invariant subvarieties given by the {\it
KdV locus condition}
\begin{equation} \label{locus}
\sum_{i\ne j}^N \wp' (q_i-q_j)=0, \quad q_i\ne q_j, \quad j=1,\dots, N .
\end{equation}
If $N$ equals a "triangular number", i.e., $N=n(n+1)/2$ for some $n$,
 and the coordinates $q_i$ satisfy the KdV locus condition, then the following
properties hold

\begin{description}
\item{1).} \textup{(\cite{Dub_Nov, AMM77})} For a certain constant $C$, the function
\begin{equation} \label{e-KdV}
U(x,t)=2 \sum_{i=1}^N \wp (x-q_i(t)) + C
\end{equation}
is an elliptic solution of the KdV equation $U_t=6 U U_x - U_{xxx}$.

\item{2).} \textup{(\cite{Smirnov1, Smirnov2})} The Krichever
spectral curve is hyperelliptic of genus $n$: there exists a
birational change of variables $(\wp(u), \mu)\rightarrow (z,w)$
such that the equation defining ${\cal G}$ takes the canonical
form
\begin{equation}\label{KdVcurve}
w^2=R_{2n+1}(z)\equiv (z-z_1)\cdots (z-z_{2n+1})
\end{equation}
with some parameters $z_k$. Up to a constant factor, the last
holomorphic differential $\omega_n= z^{n-1} dz /w$ on  ${\cal G}$
reduces to the holomorphic differential $d\wp /\wp'$ on $\cal E$.

\item{3).} \textup{(\cite{Krich})} Let
$\psi_1,\dots,\psi_{n-1},\psi_n$ be the local coordinates on
Jac$({\cal G})$ associated to the holomorphic differentials
$\omega_k= z^{k-1} dz/w$, $k=1,\dots,n$ on ${\cal G}$ and let
$\varphi_1,\dots,\varphi_n$ be the coordinates associated to the
normalized holomorphic differentials. Let $\theta(\varphi)$ be the
theta-function associated to ${\cal G}$ and $U\in {\mathbb C}^n$
be the tangent vector of the curve ${\cal G}\subset
\mbox{Jac}({\cal G})$ at $\infty\in {\cal G}$, which in the
coordinates $\psi_k$ has the form $U=(0,\dots,0,2)^T$. Then for
suitable constant vectors $V,W \in {\mathbb C}^n $, the
transcendental equation
\begin{equation} \label{ECM}
\theta [\Delta] (Ux +Vt+W )=0, \qquad q,t \in {\mathbb C},
\end{equation}
has exactly $N$ solutions $x=q_1(t), \dots, x=q_N(t)$ on Jac(${\cal G}$),
which provide the solution for the EC system.

\item{4).} \textup{(\cite{Gavr_Per})} One can choose a canonical
homology basis on ${\cal G}$ and the basis of {\it normalized}
holomorphic differentials such that the vectors $U,V$ in
(\ref{ECM}) take the form
\begin{equation} \label{norms}
U=(1,\dots,0,0)^T, \quad V=(0, V_1,\dots,V_{n-1})^T
\end{equation}
and the period matrix of $\cal G$ becomes
\begin{equation} \label{periodsN}
( 2\pi \jmath {\bf I}\; B), \quad
B= \begin{pmatrix} \tau/N & 2\pi \jmath/N & 0 & \cdots & 0 \\
          2\pi \jmath/N   &  &   &  & \\
                    0     &  &   &   & \\
                   \vdots &  &   & {\mathfrak B}  & \\
                    0     &  &      & & \end{pmatrix} ,
\end{equation}
where $( 2\pi \jmath, \tau)$ is the normalized period matrix of
the elliptic curve $\cal E$ and ${\mathfrak B}$ is a $(n-1)\times
(n-1)$ Riemann matrix.

In view of normalization (\ref{norms}), the function
$\theta[\Delta](Ux+W)$ can be written as $\theta[\Delta](x, {\bf
t} )$, ${\bf t}=(t_2,\dots,t_n)$.  It is a one-dimensional
theta-function of order $N$ with respect to $x$ and it admits the
factorization
\begin{equation} \label{theta_N}
\theta [\Delta] (x,{\bf t}) = \zeta ({\bf t}) \prod_{j=1}^N
\theta_{11}(x-q_j({\bf t})),
\end{equation}
where $\theta_{11}(x \mid \tau)\equiv \theta \left[ { 1/2 \atop
1/2} \right](x \mid \tau)$ is the one-dimensional Riemann
theta-function associated to $\cal E$ and the factor $\zeta ({\bf
t})$ depends on ${\bf t}$ only.
\end{description}

Notice that item (3) provides an implicit solution for the EC
system. Explicit expressions for certain symmetric functions of
the coordinates $q_i$ were first given in \cite{Gavr_Per}.

According to the Poincar\'e reducibility theorem (see e.g.,
\cite{BEBIM}), apart from the curve $\cal E$, the Jacobian of
$\cal G$ contains an $(n-1)$-dimensional Abelian subvariety ${\cal
A}_{n-1}$ and is isogenous to the direct product ${\cal E}\times
{\cal A}_{n-1}$.

Notice that for $n=2$, explicit algebraic expressions of the
covers and coefficients of hyperelliptic curves are known for
$N\le 8$ (see \cite{Tr}).

\paragraph{Remark 4.1.}
Property 3) also implies that the $x$-flow (or $\psi_n$-flow) on
Jac$({\cal G})$, which is tangent to ${\cal G}\subset {\mbox Jac
}({\cal G})$ at its infinity point $\infty$, intersects any
translate of the theta-divisor $\Theta\subset$ Jac($\cal G$)
precisely at $n(n+1)/2$ points (possibly with multiplicity).
\medskip

This property is similar to what we require for the linearization
of the geodesic flow on the $n$-dimensional quadric $Q$. However,
as seen from the quadratures (\ref{quad2}), the linearized flow on
Jac($\Gamma$) has a different direction: it is tangent to the
embedded hyperelliptic curve $\Gamma\subset \mbox {Jac}({\Gamma})$
not at $\infty$, but at the Weierstrass point ${\cal O}=(0,0)$.
Nevertheless, both behaviors are equivalent: if $(\beta,0)$ is a
finite Weierstrass point on $\cal G$, then by a birational
transformation
\begin{equation} \label{la-z}
\lambda=\frac{\alpha}{(z-\beta)}, \quad \mu= \frac w{(z-\beta)^{n+1}}
\end{equation}
the points $\infty$ and $(\beta,0)$ on $\cal G$ are sent to ${\cal
O}$ and $\infty$ on $\Gamma$, respectively. Then, identifying the
curves ${\cal G}$ and $\Gamma$ as well as their Jacobians, we find
that the KdV flow ($\psi_n$-flow) on Jac($\cal G$) is represented
as $u_1$-flow on Jac($\Gamma$) and vice versa.

Thus, the birational transformation (\ref{la-z}) establishes a
relation between the elliptic $N$-soliton KdV solutions
(\ref{e-KdV}) or the solutions of the $N$-body elliptic Calogero
systems satisfying the locus condition and the geodesics on a
quadric $Q$ that are linearized on Jac$(\Gamma)$. Moreover, since
the KdV solutions are periodic, {\it all such geodesics are
closed}.

Clearly, for the geodesic and the quadric (ellipsoid) to be real, certain reality
conditions on the coefficients of $\cal G$ must be satisfied.

In particular, assume that the roots $z_1,\dots,z_{2n+1}$ of
${\cal G}$ (see formula (\ref{KdVcurve})) have such values that
the above transformation with certain $\alpha$ and
$\beta=z_{2n+1}$ sends them to real {\it positive} numbers
$e_1,\dots,e_{2n}$ and the infinity. In this case one can always
choose a partition
$$
\{e_1,\dots,e_{2n}\} =\{a_1 ,\dots, a_{n+1}\}\cup \{c_1,\dots,c_{n-1}\}
$$
such that the quadratures (\ref{quad2}) describe the {\it real}
closed geodesics on the ellipsoid $Q$ that are tangent to the
confocal quadrics $Q(c_1), \dots, Q(c_{n-1})$.

\paragraph{Parametrization of the closed geodesics.}
Let the hyperelliptic curve $\Gamma$ be isomorphic to $\cal G$ via
(\ref{la-z}). Then, in view of property (4) of the EC solutions,
we choose  local coordinates $\varphi_1,\dots,\varphi_n$ on
Jac$(\Gamma)$ such that its period matrix has the form
(\ref{periodsN}) and the function $\theta[\Delta](U_1 s
+\varphi_0\mid B)$
%,  $\theta[\Delta\eta_i](U_1 s +\varphi_0\mid B)$
defined in (\ref{theta_n}), (\ref{etas}) can be written as
$\theta[\Delta](s , {\bf t}\mid B)$, ${\bf t} =(t_2,\dots,
t_n)^T$=const. Here, as above, $s $ is the new parameter
introduced in (\ref{tau-1}) and ${\bf t}$ plays the role of the
constant phase vector numerating the geodesic.

Now let
$$
\{ q_1 ({\bf t}) ,\dots,q_N({\bf t})\} \quad \textup{and } \quad
\{ p_{1i}({\bf t}),\dots,p_{Ni} ({\bf t})\},  \qquad i=1,\dots, n
$$
be solutions to equations
\begin{gather} \label{p_i}
\theta [\Delta] (s , {\bf t})=0, \quad \mbox{respectively} \quad
\theta[\Delta + \eta_{i} ](s , {\bf t})=0, \qquad s  \in {\mathbb
C} .
\end{gather}
% where the half-integer characteristics $\eta_{i}, \Delta$ ???

\begin{proposition} \label{close}
\begin{description}
\item{1).} The Cartesian coordinates $X_i$ of the closed geodesics
admit the following parametrization in terms of theta-functions of
$\cal E$:
\begin{equation} \label{periodic}
X_i (s  | {\bf t}) = \xi_i ({\bf t} ) \exp (s  /2) \cdot \frac {
\theta_{11}(s  -p_{1i}( {\bf t} ) ) \cdots \theta_{11}(s  -p_{N
i}({\bf t}) ) } { \theta_{11}(s  -q_{1}({\bf t})) \cdots
\theta_{11}(s  -q_{N}( {\bf t}) ) }\, .
\end{equation}
Here the factors $\xi_i ({\bf t})$ depend on the phases only, and,
for any ${\bf t}$, modulo the periods $(2\pi \jmath ,\tau)$ of
$\cal E$,
\begin{equation}\label{match}
p_{1i}+\cdots+p_{Ni} \equiv q_1 +\cdots+q_N+ \tau/2 .
\end{equation}

\item{2).} Let the parameter $d\notin \{a_1 ,\dots, a_{n+1},\,
c_1,\dots,c_{n-1}\}$ be such that the real closed geodesics on $Q$
have a non-empty intersection with the quadric $Q_d$. Then the
geodesics intersect each connected component ${D}_{d,j}$ of $Q\cap
Q_d$ at least at 2 and at most at $n(n+1)$ distinct points.
\end{description}
\end{proposition}

As follows from (\ref{periodic}), (\ref{match}), as well as the
quasi-periodic law of $\theta_{11}(s )$, when the argument $s $
changes by the period $2\pi \jmath$ of $\cal E$, the coordinates
$X_i$ change sign, and they remain unchanged under the shift $s
\to s +\tau$. Hence the corresponding closed geodesic
is a 2-fold unramified covering of $\cal E$, % obtained by doubling its periods
i.e., it is an algebraic curve in ${\mathbb R}^{n+1}$.
\medskip

\noindent{\it Proof of Proposition} \ref{close}. 1). Due to definition of generic
theta-functions with characteristics (\ref{char}), one has
$$
\theta[\Delta + \eta_{i} ](\varphi) =\exp\{\langle \eta_i',
B\eta_i' \rangle +\langle
\varphi+2\pi\jmath(\eta_i''+\Delta_i''),\eta_i'\rangle \}\,
\theta[\Delta](\varphi+2\pi \jmath\, \eta_i''+B \eta_i')
$$
and therefore,
\begin{align}  \theta[\Delta + \eta_{i} ](s , {\bf t})
& = \varkappa_i({\bf t}) \,\exp (s  (\eta_i)_1' ) \nonumber \\
&\cdot \theta[\Delta]\left( s  + \tau \, \frac{ (\eta_i)_1' }{N} +
2\pi\jmath\left[\frac{(\eta_i)_2'}{N}+ (\eta_i)_1''\right], {\bf
t}+ {\mathfrak t} \right) \label{displace}
\end{align}
with certain factors $\varkappa_i({\bf t})$ and vector ${\mathfrak
t}\in {\mathbb C}^{n-1}$. One can show that, for this choice of
the basis of cycles and of the normalized differentials on
$\Gamma$, one always has $(\eta_i)_1'=1/2$. Next, in view of
(\ref{p_i}) and (\ref{theta_N}), the theta-function in the right
hand side of (\ref{displace}) admits factorization in
one-dimensional theta-functions $\theta_{11}(s - p_{ki}\mid
\tau)$. Now applying this to the solution (\ref{theta_n}) for the
Cartesian coordinates of a generic geodesic, one arrives at
expressions (\ref{periodic}).

Since the geodesics are closed, these expressions must be doubly periodic in $s $.
Then, due to the quasiperiodic law for $\theta_{11}(s )$, relation (\ref{match})
must hold.
\medskip

\noindent 2).
As shown in Section 2, under the Abel--Jacobi map (\ref{AB})
the intersections of the geodesic with $Q \cap Q_d$
correspond to points on the two translates
$\Theta_{d-}, \Theta_{d+}$ of the theta-divisor on Jac($\Gamma$).
Next, according to property 3) of the EC solutions, the $u_1$-flow on Jac($\Gamma$)
intersects any translate of $\Theta$ at $N=n(n+1)/2$ (complex) points.

According to Remark 2.1., the geodesic flow is linearized on a
covering $\widetilde{\rm Jac}(\Gamma)$ obtained by doubling of
some of its periods. As a result, the $u_1$-flow on
$\widetilde{\rm Jac}(\Gamma)$ has in total $4N=2n(n+1)$
intersections with the corresponding coverings of $\Theta_{d-}\cup
\Theta_{d+}$, which gives rise to the same number of complex
intersections of the geodesic with $Q\cap Q_d$.

For $d\ne a_i$, the real manifold $Q\cap Q_d$ consists of 2
connected components $D_{d,1}, D_{d,2}$. Then, due to symmetry,
each of the components is intersected at most at $n(n+1)$ points.
 \boxed{}
\medskip

\paragraph{The case $N=3$, $n=2$.}
In this simplest case the connected components $D_{1,d}, D_{2,d}$
are ovals which bound simply-connected domains ${\cal D}_{1,d},
{\cal D} _{2,d}$ on the 2-dimensional ellipsoid $Q$. Then one can
define {\it ingoing} and  {\it outgoing} intersections of the
geodesic with $D_{i,d}$ (when the geodesic goes inside,
respectively outside of ${\cal D}_{i,d}$). In view of Proposition
\ref{main0}, the former ones correspond to the intersections of
the $u_1$-flow on Jac($\Gamma$) with $\Theta_{d+}$, whereas the
latter correspond to the intersections with the other translate
$\Theta_{d-}$.

As follows from item (2) of Theorem \ref{close}, the closed
geodesic has 3 ingoing and 3 outgoing complex intersections with
each component $D_{i,d}$. If $d\to a_i$, both components are
confluent to the ellipse $Q\cap \{X_i=0\}$ and, therefore, the
closed geodesics intersect the hyperplane $X_i=0$ at 6 (generally
complex) points. Since the geodesic is an elliptic curve in
${\mathbb C}^3$, it can be represented as an intersection of the
quadric $Q$ with a cubic surface (\cite{Per}).

The tangential 3:1 covering ${\cal G}\mapsto {\cal E}$ is
associated to 3-elliptic KdV solutions and dates back to works of
Hermite and Halphen (\cite{Herm}). The genus 2 curve ${\cal G}$ is
birationally equivalent to
\begin{equation} \label{3:1}
H =\left\{w^2 = -\frac 14 (4 z^3-9 g_2 z- 27 g_3 ) (z^2 -3g_2) \right \} \,
\end{equation}
which covers the elliptic curve with moduli $g_2, g_3$,
$$
{\cal E}_1 =\{  W^2 = 4 Z^3 - g_2 Z -g_3 \}\subset (Z,W) .
$$
The covering is given by the relations
$$
Z=-\frac 19 \frac{z^3-27 g_3}{z^2-3g_2}, \quad
W=\frac 2{27} \frac {w(z^3 -9g_2 z+ 54 g_3)}{(z^2-3g_2)^2} ,
$$
and possesses 2 branch points.
The last (second) holomorphic differential on $H$ reduces to an elliptic one:
$$
\frac {z\, d z}{w} =- \frac 23 \frac {d Z}{W} .
$$
%This cover corresponds to special solutions of the three body elliptic Calogero system,
%when $q_1, q_2, q_3$ satisfy the locus condition (\ref{locus}), and to 3 elliptic
%KdV solution.

In this case there is another 3:1 covering $\Pi_2\; :\; H\mapsto {\cal E}_2$
of the elliptic curve
\begin{gather}
{\cal E}_2 =\{ {\cal W}^2= {\cal Z}^3 -G_2 {\cal Z}-G_3  \}, \label{E2} \\
 G_2=\frac {27}{16}(g_2^3+9g_3^2), \quad G_3=\frac {243}{32}(3g_3^2-g_2^3), \label{G}
\end{gather}
such that
\begin{equation} \label{2nd-cover}
{\cal Z}=-\frac 14 (4z^3-9 g_2 z-9g_3),\quad {\cal W}=- w\left(
z^2-\frac 34 g_2\right), \quad \mbox{and } \quad
 \frac {d z}{w} = \frac 1 3 \frac {d {\cal Z}}{\cal W} .
\end{equation}

Let $(u_1, u_2)$ be local coordinates on Jac$(H)$ associated to
the holomorphic differentials $(k_1 \,z\,dz/w, k_2\, dz/w)$,
$k_1,k_2=$const. According to property (4) of the elliptic KdV
solutions, for suitable normalization constants $k_1,k_2$ the
period matrix of Jac$(H)$ takes the form
\begin{equation}\label{periods}
\begin{pmatrix} 2\pi \jmath & 0 & \tau /3 &  2 \pi \jmath/3 \\
                     0  & 2\pi \jmath & 2 \pi \jmath/3 & {\mathfrak b} /3
\end{pmatrix} \, ,
\end{equation}
where $(2\pi\jmath, \tau)$ and $(2\pi\jmath, {\mathfrak b})$ are
the normalized period matrices of ${\cal E}_1$ and ${\cal E}_2$
respectively.

\paragraph{Algebraic conditions for the closed geodesics.}
Under the birational equivalence of the  curves $H$ and $\Gamma$,
the structure of $H$ imposes conditions on the parameters $\{a_1,
a_2, a_3, c_1\}=\{e_1,\dots, e_4\}$ (squares of semi-axes and the
constant of motion), for which the geodesics are closed, and which
can be found explicitly. Notice that one can always set
$e_1=a_1=1$ choosing appropriately $\alpha$ in (\ref{la-z}). Then
$\infty\in \Gamma$ is transformed  either to one of roots
$\pm\sqrt{3g_2}$ or to one of the roots of $4z^3-9g_2z-g_3$ in
(\ref{3:1}). In the first case $e_2,e_3,e_4$ are positive, hence
$Q$ is an ellipsoid, in the other case some of $e_i$ are negative,
so $Q$ is a hyperboloid. In the latter case, reality condition
implies that is the geodesic flow on the hyperboloid corresponds
to unbounded motions, which we do not consider here.

These observations lead to the following sufficient conditions for
a real geodesic on $Q$ to be closed.

\begin{proposition} \label{e1-e4}
For given parameters $e_1=1$, $e_2>0$ the geodesic associated to the covering
$H \to {\cal E}$ is closed if and only if one of the following pairs of conditions on
$e_3$ and $e_4$ is satisfied
\begin{equation}\label{condB}
\left(1+\frac{1}{e_2}+\frac{1}{e_3}\right)^2
-4\left(\frac{1}{e_2}+\frac{1}{e_3}+\frac{1}{e_2e_3}\right)=0, \quad\quad
e_4=\frac{3e_2e_3}{2(e_2+e_3+e_2e_3)},
\end{equation} or
\begin{equation}\label{condC}
9e_2^2e_3^2-24e_2e_3(e_2+e_3)+16 e_2e_3+16 e_2^2+e_3^2=0,\quad\quad
e_4=\frac{2e_2e_3}{3e_2e_3- 2(e_2+e_3)}.
\end{equation}
If (\ref{condB}) is satisfied, then the following ordering holds
\begin{align*}
\textup{for   } \quad 0<e_2<\frac14, \quad &0<e_4<e_2<e_3<1, \\
\textup{for   } \quad  \frac14<e_2<1 \quad &0<e_4<e_2<1<e_3, \\
\textup{for   } \quad  1<e_2<4, \quad &0<e_4<1<e_2<e_3, \\
\textup{for   } \qquad  e_2>4, \quad & 0<e_4<1<e_3<e_2.
\end{align*}
Condition (\ref{condC}) is satisfied if and only if $e_2>1$, and the following
ordering of roots holds:
\begin{align*}
\textup{for   } \quad  1<e_2<\frac43, \quad & 0<1<e_2<e_3<e_4 \\
\textup{for   } \quad  \frac43<e_2<4, \quad & 0<1<e_3<e_2<e_4, \\
\textup{for   }  \qquad  e_2>4, \quad & 0<1<e_3<e_4<e_2.
\end{align*}
\end{proposition}

\paragraph{Remark 4.2.} The first condition corresponds to the case when $e_1=1$ is
transformed into one of the roots of the polynomial $4z^3-9g_2z-g_3$, while in the
second case $1$ is transformed to $\sqrt{3g_2}$.
\medskip

The inverse relations are described by the following lemma.

\begin{lemma}\label{betas} If $e_2, e_3, e_4$ is a solution of (\ref{condB}), then
$$
\beta = -\frac 13\left (1+ \frac 1{e_2} +\frac 1{e_3} \right),
\quad g_2= \frac 13 \beta^2, \quad g_3=-\frac
4{27}(\beta+1)\left(\beta+ \frac 1{e_2}\right )\left(\beta+ \frac
1{e_3}\right) .
$$
If condition (\ref{condC}) is satisfied, then
$$
\beta = -\frac 12, \quad g_2= \frac 1{12}, \quad
g_3=-\frac 1{27} \left(\frac 18-\frac 1{e_2 e_3 e_4} \right).
$$
\end{lemma}

Proposition \ref{e1-e4} and Lemma \ref{betas} are proved by pure calculations.

% These cases lead to different sufficient conditions for the geodesic to be closed.

\paragraph{Remark 4.3.}
Since Proposition \ref{e1-e4} imposes two independent conditions
on the parameters $a_2, a_3, c$, not any triaxial ellipsoid has
closed geodesics associated to the 3:1 tangential covering $H \to
{\cal E}$. Naturally, this does not exclude existence of closed
geodesics for the other types of hyperelliptic tangential coverings
considered in \cite{TV, Tr}. Description of such geodesics will be
the subject of a future study.

\section{Periodic billiard trajectories on a 2-dimensional quadric}
As shown in Section 2, a criterium for a generic billiard on an
ellipsoid $Q$ to be periodic cannot be established in the complex
setting. Below we restrict ourselves to
2-dimensional ellipsoid $Q$ for which the trajectories between impacts
on $Q\cap Q_d$ are segments of the closed geodesics $X(s)$ described
by Proposition \ref{e1-e4}. We study the following problem: given
a family of such closed geodesics with an appropriate fixed caustic parameter
$c$, for which boundary parameters $d$ the corresponding billiard
orbit is periodic?

\paragraph{Correctness of the problem.}
% The above problem requires a further clarification:
As stated in the previous section, the closed geodesic $X(s)$ has
3 ingoing and 3 outgoing complex intersections with each component
$D_{i,d}$ of $Q\cap Q_d$. Hence, to an initial set $(x,v)$, $x\in
D_{i,d}$ there correspond 3 possible branches $(\tilde x, \tilde
v)$, i.e., the restricted complex map ${\cal B}_c$ is 3-valued. In
this context the problem of its periodicity makes no sense, unless
other properties are taken into account.

In the real domain there may exist one, two, or three real
branches $\{(\tilde x, \tilde v)\}$, and one naturally chooses a
unique set $(\tilde x^*, \tilde v^*)\in \{(\tilde x, \tilde v)\}$,
such that $x$ and $\tilde x^*$ are subsequent points of
intersection of $X(s)$ with $D_{i,d}$. However, $(\tilde x^*,
\tilde v^*)$ cannot be distinguished form the other branches by
purely algebraic methods. Yet the 3 branches have an important
common property.

\begin{lemma}
Given an initial set $(x,v)$, the 3 sets $(\tilde x, \tilde v)$ generate
one and the same closed geodesic $\tilde X(s)$ on $Q$.
\end{lemma}

Indeed, according to Theorem  \ref{main1}, under the billiard map
${\cal B}_c$ the local coordinate $u_2$ on Jac$(\Gamma)$ increases
by the integral ${\mathfrak q}_2=\int_{E_{d-}}^{E_{d+}}
\frac{\lambda d\lambda}{\mu}$, which depends on $d$ only. Hence,
the images of the sets $(\tilde x, \tilde v)$ in Jac$(\Gamma)$
have the same coordinate $u_2$ and, therefore, they belong to the same
trajectory of the $u_1$-flow on Jac$(\Gamma)$. \boxed{}
\medskip

It follows that the map  ${\cal B}_{c}$ induces a map on the set of the closed geodesics,
${\cal F}_{c,d}\, : X(s)\mapsto \widetilde X(s)$, which is one-to-one.

If the orbit of ${\cal F}_{c,d}$ is periodic, then the orbit of
the transcendental real one-to-one map ${\cal B}_c^*\, : (x,v)\to (\tilde x^*, \tilde
v^*)$ is periodic as well. Hence, an algebraic condition of
periodicity for ${\cal F}_{c,d}$ gives us a correctly defined
periodicity condition for the billiard on $Q$.

\paragraph{Periodicity conditions for  ${\cal F}_{c,d}$.} The map ${\cal F}_{c,d}$
can be described as the shift $u_2 \mapsto u_2 +{\mathfrak q}_2$
and under the birational transformation (\ref{la-z}) the increment
${\mathfrak q}_2$ takes the form
$$
{\mathfrak q}_2=\int_{E_{\delta -}}^{E_{\delta +}} \frac{d z}{w},
\qquad \delta=1/d+\beta, \quad E_{\delta \pm}=(\delta,
\pm\sqrt{R(\delta)})\in H .
$$
Next, under the second covering $\Pi_2\; :\; H\mapsto {\cal
E}_2\subset ({\cal Z}, {\cal W})$ described by (\ref{2nd-cover}),
the above integral reduces to the elliptic one: ${\mathfrak
q}_2=\frac 13{\cal I}_\rho$,
$$
{\cal I}_\rho= \int^{ \Pi_2(E_{\delta +}) }_{\Pi_2(E_{\delta -})} \,
\frac {d {\cal Z}}{\sqrt{ {\cal Z}^3 -G_2 {\cal Z}-G_3}} , \qquad
 \Pi_2( E_{\delta \pm }  )=(\rho, \pm\sqrt{\rho^3 -G_2 \rho-G_3})\, ,
$$
where
\begin{equation} \label{rho}
\rho =-(\delta^3-9/4 g_2 \delta-9/4 g_3).
\end{equation}

As follows from the structure of the period matrix of ${\cal G}$
in (\ref{periods}), the shift $u_2\mapsto u_2 +{\mathfrak q}_2$
results in the same $u_1$-winding on Jac$(\Gamma)$ if ${\mathfrak
q}_2$ belongs to the lattice $\Lambda_0= \{2\pi\jmath {\mathbb Z}
+ \frac 13 {\mathfrak b} {\mathbb Z}\}$. Hence, the orbit of
${\cal F}_{c,d}$ is periodic if and only if ${\mathfrak q}_2$ is a
finite order point of $\Lambda_0$.

Since the period lattice $\Lambda=\{2\pi\jmath {\mathbb Z} +{\mathfrak b}{\mathbb Z}\}$
of Jac$({\cal E}_2)$ is a sublattice of $\Lambda_0$, we conclude
that ${\cal F}_{c,d}$ is periodic if, in particular, for an integer $m$,
the integral $m {\cal I}_\rho$ is neutral in Jac$({\cal E}_2)$.
(Notice that one should exclude trivial cases, when ${\cal I}_\rho$ is neutral in
the lattice $\Lambda_0$.)

Then, by applying the Cayley-type condition (Theorem \ref{Drag0}),
one obtains an algebraic equation on the parameter $\rho$. %, which we  obtain below.

In practice, for large $m$, such equations appear to be rather tedious.
We consider only the simplest nontrivial case $g=1, m=3$, when the condition (\ref{rank})
for the elliptic curve ${\cal E}_2$ takes the form
\begin{equation} \label{N3}
\bigg |\begin{matrix}
S_{4} & S_{3} \\
S_{5} & S_{4}
\end{matrix} \bigg | =0,
\end{equation}
$S_j$ being the coefficients of the expansion
\begin{align*}
\sqrt{{\cal Z}^{3}-G_{2} {\cal Z}-G_{3}}&\equiv\sqrt{
({\cal Z}-\rho)^{3}+P_{2}({\cal Z}-\rho)^{2}+P_{1}({\cal Z}-\rho)+P_{0}} \\
& = S_0 + S_1 ({\cal Z}-\rho)+ S_2 ({\cal Z}-\rho)^2+ \cdots ,
\end{align*}
where
\begin{equation} \label{P3}
P_0  =  \rho^3 -G_2 \rho -G_3, \quad P_1  = 3 \rho^2 -G_2 , \quad
P_2  =  3\rho.
\end{equation}
Condition (\ref{N3}) yields the equation
\begin{align*}
 & ( 4P_{0}P_{2}-P_{1}^{2}) \\
 & \; \cdot
\left (512P_{0}^{4}+3P_{1}^{6}-384P_{0}^{3}P_{1}P_{2}-20P_{0}P_{1}^{4}P_{2}
+96P_{0}^{2}P_{1}^{3} +64P_{0}^{3}P_{2}^{3}+16P_{0}^{2}P_{1}^{2}P_{2}^{2} \right)=0.
\end{align*}
Substituting here expressions (\ref{P3}), we conclude that the
integral $m \,{\cal I}$ is neutral in Jac$({\cal E}_2)$ if $\rho$
is a root of one of the following two equations
\begin{align}
& \qquad 3\rho^4-6 G_2 \rho^2 -12 G_3 \rho -G_2^2=0 , \label{cond3} \\
 & -\rho^{12}+22 G_{2}\, \rho^{10} +220 G_{3}\, \rho^{9} +165G_{2}^{2}\, \rho^{8}
+ 528 G_{2}G_{3}\, \rho^{7} \nonumber \\
& +\left( 1776G_{3}^{2}-92G_{2}^{3}\right)\, \rho^{6} +264 G_{2}^{2}G_{3}\, \rho^{5}
+ \left( 185G_{2}^{4}-960G_{2}G_{3}^{2}\right) \rho^{4} \nonumber \\
& + \left( 80G_{2}^{3}G_{3}-320G_{3}^{3}\right) \rho^{3}
  + \left( 624G_{2}^{2}G_{3}^{2}-90G_{2}^{5}\right) \rho^{2}
+\left( 896G_{2}G_{3}^{3}-132G_{2}^{4}G_{3}\right) \rho \nonumber \\
 & +512G_{3}^{4}+3G_{2}^{6}-96G_{2}^{3}G_{3}^{2}=0 . \label{cond4}
\end{align}

\paragraph{Remark 5.1.}
The fact that equation (\ref{N3}) has 16 roots $\{\rho_j\}$ is
natural: they correspond to 32 points $\left\{ E_{\rho_j \pm}
=\left(\rho_j,\pm\sqrt{\rho_j^3 -G_2 \rho_j-G_3}\right )\right\}$
on the curve ${\cal E}_2$ that satisfy conditions
$$
E_{\rho_j +}\ne E_{\rho_j -}, \quad 6 \, {\cal A}(E_{\rho_j\pm })
\equiv 0, \qquad \mbox{where  }\quad {\cal A}(P) =\int_{\cal
O}^{P} \frac {d {\cal Z}}{\cal W},
$$
$\cal O$ being a Weierstrass point on ${\cal E}_2$. The
corresponding 32 points ${\cal A}(E_{\rho_j\pm })
\in\mbox{Jac}({\cal E}_2)$ form a complete set of sixth order
points $\{( {\mathbb Z}/6{\mathbb Z}  ) 2 \pi\, \jmath + ({\mathbb
Z} /6{\mathbb Z} ){\mathfrak b}/2\}$ of which the four second
order points $\{0, \pi\, \jmath, {\mathfrak b}/2, \pi\,
\jmath+{\mathfrak b}/2\}$ are excluded.
\medskip

The resulting algorithm of constructing real periodic billiard
trajectories on $Q$ consists of 4 steps.
\begin{description}
\item{1)} Given a family of closed geodesics on $Q$ defined by
parameters $e_1=1, e_2, e_3, e_4$ in Proposition 3.2, one finds
the constant $\beta$ in (\ref{la-z}) and moduli $g_2, g_3$ of the
first elliptic curve ${\cal E}_1$ by using Lemma \ref{betas} and
then, in view of (\ref{G}), the parameters $G_2, G_3$ of the curve
${\cal E}_2$.

\item{2)} One finds the set $\{\rho_j \}$ of all roots of
equations (\ref{cond3}), (\ref{cond4}) and then rejects the
"trivial" ones, for which the integral ${\cal I}_\rho$ is neutral
in the lattice $\Lambda_0$.

\item{3)} One computes the sets of corresponding admissible
parameters $\delta$ and $d$ by solving the cubic equation
(\ref{rho}) and the linear equation $\delta=1/d+\beta$.

\item{4)} One selects the subset of real positive admissible
values of $d$ for which the quadric $Q_d\subset {\mathbb R}^3$ has
a nonempty intersection with the closed geodesics on $Q$, that is,
the values that belong to segments $\{-\lambda
(\lambda-e_1)\cdots(\lambda-e_4)>0\}$.
\end{description}

The parameters $e_1=1, e_2, e_3, e_4$ and the selected real values of $d$ characterize
the billiard boundary $Q\cap Q_d$ and the caustics
for which the nontrivial billiard orbits are periodic.

%%%%%%%%%%%%%%%%%%%%%%%%%%%%%%%%%%%%%%%%%%%%%%%%%%%%%%%%%%%%%

\section*{Conclusion} In this paper we studied
algebraic geometrical properties of the billiard on a quadric $Q$
with bounces along its intersection with the confocal quadric
$Q_d$.

If $Q$ is a 2-dimensional ellipsoid, we explicitly described
a special class of closed geodesics associated to the 3-soliton
solutions of the KdV equation corresponding to the 3-fold tangential covering
${\cal G}\mapsto {\cal E}_1$.

The existence of the second elliptic curve ${\cal E}_2\subset
\mbox{Jac} ({\cal G})$ in this case enabled us to apply the
Cayley-type periodicity condition (Theorem \ref{Drag0}) as an
algebraic description of finite order points on ${\cal E}_2$ and,
thereby, to obtain an explicit description of 3-periodic billiard
orbits on $Q\cap Q_d$ whose points are joined by arches of the
closed geodesics. As mentioned above, for period $m>3$ the
periodicity conditions lead to quite tedious algebraic equations
for the boundary parameter.

This approach cannot be directly extended to construct
similar-type periodic billiard orbits on an $n$-dimensional
quadric $Q$, since, instead of the complimentary elliptic curve
${\cal E}_2$, the Jacobian of ${\cal G}$ contains an
$(n-1)$-dimensional Prym subvariety ${\cal A}_{n-1}$, which, in
general, is not a Jacobian variety or its covering (except $n=3$).

Thus, we arrive at the problem of an algebraic description of finite-order points in
${\cal A}_{n-1}$ in terms of points of the curve ${\cal G}$, which deserves a separate
studying.

\subsection*{Acknowledgments}
We thank E.Previato, V. Dragovi\'c, and A.Treibich for useful
discussions, as well as A. Perelomov for valuable remarks during
preparation of the manuscript and indicating us the reference
\cite{Braun}.

The second author (Yu.F.) acknowledges the support of grant BFM 2003-09504-C02-02 of
Spanish Ministry of Science and Technology.

This research was partially supported by GNFM-INdAM project ``Onde nonlineari,
struttura tau e geometria delle varietà invarianti: il caso della gerarchia di
Camassa-Holm''.

\end{document}